\documentclass[12pt,fleqn]{article}
\usepackage{psfig,float}
\usepackage{graphicx}
\usepackage{bm}

\begin{document}

\begin{center}
{\Large{\bf $\eta \to \pi^0 \gamma \gamma$ decay within a 
chiral unitary approach }}
\end{center}
\vspace{1cm}

\title{$\eta \to \pi^0 \gamma \gamma$ decay within the chiral
unitary approach. }

\begin{center}
{\large{E. Oset$^1$, J. R. Pel\'aez$^2$ and L. Roca$^1$}}

 \vspace{.5cm}
{\it $^1$ Departamento de F\'{\i}sica Te\'orica and IFIC
 Centro Mixto Universidad de Valencia-CSIC\\
 Institutos de Investigaci\'on de Paterna, Apdo. correos 22085,
 46071, Valencia, Spain}    

\vspace{.5cm}{\it $^2$ Dip. di Fisica. Universita' degli Studi, Firenze, and INFN, Sezione di Firenze, Italy\\
   Departamento de
F\'{\i}sica Te\'orica II,  Universidad Complutense. 28040 Madrid,
Spain.
}  
\end{center}

\begin{abstract}
We improve the calculations of the 
$\eta \to \pi^0 \gamma \gamma$ decay within
the context of meson chiral lagrangians. We use a chiral unitary
approach for the meson-meson  interaction,
thus generating the $a_0(980)$ resonance and fixing
the longstanding sign ambiguity on its contribution.
This also allows us to calculate the loops
with one vector meson exchange, thus removing a former source of uncertainty. 
In addition we ensure the consistency of the approach
with other processes. First, by
using vector meson dominance
couplings normalized to agree with radiative vector meson decays.
And, second, by checking the consistency of the calculations with the
related $\gamma \gamma \to \pi^0 \eta$ reaction.
We find an $\eta \to \pi^0 \gamma \gamma$ decay width of $0.47\pm 0.10$ eV, 
in clear disagreement with published data but
in remarkable agreement with the most recent measurement.
\end{abstract}

\section{Introduction}

The $\eta \to \pi^0 \gamma \gamma$ decay has attracted much theoretical 
attention, since Chiral Perturbation Theory (ChPT) calculations
have sizable uncertainties and 
produce systematically rates 
about a factor of two smaller than experiment \cite{exp,Hagiwara:pw}. In contrast
models using quark box
diagrams \cite{Ng:sc,Nemoto:1996bh} claim to obtain acceptable rates.
Within ChPT, the problem stems from the fact that the
tree level amplitudes, both at $O(p^2)$ and $O(p^4)$, vanish.
The first
non-vanishing contribution comes at $O(p^4)$, but either from loops involving
kaons, 
largely suppressed due to the kaon masses, or from pion loops, again
suppressed since they violate G parity and are thus
 proportional to $m_u -m_d$ \cite{Ametller:1991dp}. 
The first sizable contribution comes at 
$O(p^6)$ but the coefficients involved are not precisely determined. One must
recur to models: either Vector Meson Dominance (VMD) 
\cite{Ametller:1991dp,oneda,Picciotto:sn}, the 
Nambu-Jona-Lasinio model
(NJL) \cite{Bel'kov:1995fj}, or the extended Nambu-Jona-Lasinio model
(ENJL) \cite{Bellucci:1995ay,Bijnens:1995vg}, have
been used to determine these parameters.
However, the use of tree level VMD 
to obtain the $O(p^6)$ chiral coefficients
by expanding the vector meson propagators, leads \cite{Ametller:1991dp}
to results about a factor 
of two smaller than the "all order" VMD term, which means keeping
the full vector meson propagator.
All this said, the chiral approach has been useful 
to unveil the physical mechanisms responsible for this decay,
but it has become clear that
the strict chiral counting has to be abandoned since the $O(p^6)$ 
and higher orders involved in the full (``all order'') VMD results
are larger than those of $O(p^4)$.  
For a review, see \cite{Achasov:2001qm}, together with an 
experimental upper bound.

Once the ``all order'' VMD results is accepted as the dominant mechanism,
one cannot forget the tree level exchange of other resonances
around the 1 GeV region. 
In comparison with VMD, the 
exchange of $J^{PC}=1^{+-}$ axial vectors
\cite{Ko:1992zr,Ko:rg} yields negligible contributions
when using 
values of the couplings in agreement
with $\gamma \gamma\to \pi^0 \pi^0$ data \cite{Jetter:1995js}. 
Still at tree level, the  $a_0(980)$
exchange, which was
taken into account approximately in \cite{Ametller:1991dp}, 
was one of the main sources of uncertainty, since
even the sign
of its contribution was unknown.

After the tree level light resonance exchange have been 
taken into account, we should consider loop diagrams,
since meson-meson interaction or rescattering can be rather strong. First
of all we find the already commented $O(p^4)$ kaon loops from ChPT,
but also the meson loops from the terms involving
the exchange
of one resonance. The uncertainty from the latter was roughly 
expected \cite{Ametller:1991dp} to be about 30\% of the full width.

 Another relevant question is that no attempts have been done to
check the consistency of $\eta \to \pi^0 \gamma \gamma$ results
with data from other processes.
On the one hand, the decay results have not been compared
 with the crossed
channel $\gamma \gamma\to \pi^0 \eta$, although
some consistency tests with $\gamma \gamma\to \pi^0 \pi^0$
have been carried out as quoted above. 
The reason is not surprising since there are no
hopes within ChPT to reach the $a_0(980)$ region where
there are measurements of the $\gamma \gamma\to \pi^0 \eta$ cross section 
\cite{Oest:1990ki,Antreasyan:1985wx}.
On the other hand, the explicit SU(3) breaking already present in
the 
radiative vector meson decays has not been taken
into account when calculating the VMD tree level contributions.

The former discussion has set the stage of the problem and the 
remaining uncertainties that allow for further improvement.
In recent years, with the
advent of unitarization methods, 
it has been possible to extend the results of 
ChPT to higher energies where  the perturbative expansion
breaks  down and to generate resonances up to 1.2 GeV
\cite{Oller:1997ti,Kaiser:fi,OllOsePel,Oller:1999zr,Nieves:2000bx,GomezNicola:2001as}. 
In particular these
ideas were used to describe the $\gamma \gamma \to meson-meson$ reaction, with
good results in all the channels up to energies of around 1.2 GeV 
\cite{Oller:1997yg}. Work in a
similar direction for this latter reaction has also been done in 
\cite{Dobado:1992zs,Yamagishi:1995kr,Lee:1998mz}. 
With these techniques,
and always within the context of meson chiral lagrangians,
 we will address three of the problems stated above:
First,  the $a_0(980)$ contribution, second,  the evaluation of meson loops
from VMD diagrams and, third, the consistency with the  
crossed channel $\gamma \gamma\to \pi^0 \eta$.
In particular, we will make use of the results
in  \cite{Oller:1997yg}, where the $\gamma \gamma\to \pi^0 \eta$
cross section around the $a_0(980)$ resonance  
was well reproduced  using the 
same input as in 
meson meson scattering, without introducing any extra parameters.

With these improvements we are then left with a model
that includes the ``all order'' VMD and resummed chiral loops.
We expect this approach to provide a good
description of  $\eta \to \pi^0 \gamma \gamma$ because  
recent studies on the
vector meson decay into two pseudoscalar mesons and one photon
\cite{Bramon:2001un,Palomar:2001vg,Marco:1999df}
 indicate that 
such a  combination of "all order" 
VMD  contribution plus the unitary summation of the
chiral loops  leads to good agreement with data in a variety of
reactions. These include 
$\phi \to \pi^0 \pi^0 \gamma$ \cite{Marco:1999df},
where the chiral loops are dominant, 
$\omega \to \pi^0 \pi^0 \gamma$ \cite{Bramon:2001un,Palomar:2001vg},
where  the VMD mechanism is dominant, 
and 
$\rho \to \pi^0 \pi^0 \gamma$ \cite{Bramon:2001un,Palomar:2001vg}
where both 
mechanisms have about the same
strength.

Concerning the fourth issue of
the SU(3) breaking present in radiative vector meson decays,
we will take it into account here by using effective couplings
normalized to reproduce the most recent experimental data.

Incidentally, 
there are preliminary results from a very recent experiment \cite{nefkens}, 
which give a decay width
about half of the previous one. 
In that work the authors refined the background subtraction,
 which was known to be rather problematic. Let us remark that, 
in view of the former
discussion,
 revisiting the previous theoretical
calculations is mandatory regardless of whether these new experimental results 
are confirmed or not.

In what follows we will address all these theoretical 
issues in detail, including
an updated estimation of the uncertainties in the calculation.
In particular, we will take into account
the experimental errors in the
radiative vector meson decays, which were neglected before,
although they will turn out to be the largest source of uncertainty.

\section{VMD contribution}

Following \cite{Ametller:1991dp} we consider the VMD
mechanism of Fig.~1
\begin{figure}[h]
\centerline{\hbox{\psfig{file=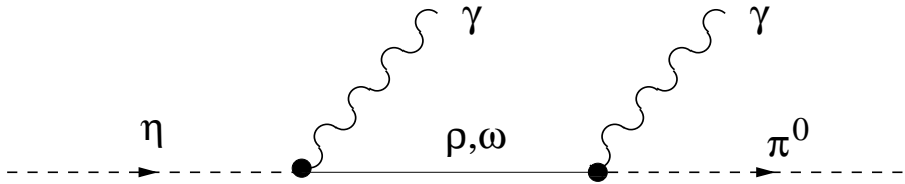,width=12cm}}}
\caption{\rm 
Diagrams for the VMD mechanism.}
\end{figure}
which can  be easily derived from the VMD Lagrangians involving
VVP and $V\gamma$ couplings \cite{Bramon:1992kr}
\begin{equation}
{\cal L}_{VVP} = \frac{G}{\sqrt{2}}\epsilon^{\mu \nu \alpha \beta}\langle
\partial_{\mu} V_{\nu} \partial_{\alpha} V_{\beta} P \rangle, \qquad
{\cal L}_{V \gamma} =-4f^{2}egA_{\mu}\langle QV^{\mu}\rangle,
\label{lagr}
\end{equation}
where $V_{\mu}$ and $P$ are standard $SU(3)$ matrices
constructed with the nonet of vector mesons containing the
$\rho$, and the nonet of pseudoscalar mesons containing the
$\pi$, respectively. For instance for pseudoscalar mesons
$P=\tilde{P}+\frac{1}{\sqrt{3}}\eta_1$, with $\tilde{P}$ given 
by \cite{Gasser:1983yg}
\begin{equation}
\tilde{P}=\left( \begin{array}{ccc}
\frac{1}{\sqrt{2}}\pi^0+\frac{1}{\sqrt{6}}\eta_8 & \pi^+  & K^+\\
\pi^- & -\frac{1}{\sqrt{2}}\pi^0+\frac{1}{\sqrt{6}}\eta_8 & K^0\\
K^- & \overline{K}^0 & -\frac{2}{\sqrt{6}}\eta_8 \\
\end{array}\right) ,
\end{equation}
and similarly for vector mesons \cite{Ecker:1988te}.
 We also assume the ordinary mixing for the $\phi$,
the $\omega$, the $\eta$ and $\eta'$:
\begin{eqnarray}
\omega &=& \sqrt{\frac{2}{3}}\omega_1 + \sqrt{\frac{1}{3}}\omega_8, \quad
\qquad 
\phi = \sqrt{\frac{1}{3}}\omega_1   - \sqrt{\frac{2}{3}}\omega_8,\\
\nonumber
\eta &=& \frac{1}{3}\eta_1 + \frac{2\sqrt{2}}{3}\eta_8, \quad
\qquad
\eta' = \frac{2\sqrt{2}}{3}\eta_1 - \frac{1}{3}\eta_8.
\label{eq:mezclas}
\end{eqnarray}
In Eq.~(\ref{lagr}) $G=\frac{3g^2}{4\pi^2f}$,
$g=-\frac{G_VM_{\rho}}{\sqrt{2}f^2}$ \cite{Bramon:1992kr}
and $f=93\,$MeV,
 with $G_V$ the
coupling of $\rho$ to $\pi\pi$ in the normalization of
\cite{Ecker:1988te}.
>From Eq.~(\ref{lagr}) one can obtain the radiative widths for
$V\to P\gamma$, which are given by
\begin{equation}
\Gamma_{V \to P\gamma}=\frac{3}{2} \alpha
 C_i^2 \left(G\frac{2}{3}\frac{G_V}{M_V}\right)^2k^3,
\end{equation}
where $k$ is the photon momentum for the vector meson at rest and
$C_i$ are $SU(3)$ coefficients that we give in Table 1 
 for the different radiative decays, together
with the theoretical (using $G_V=69\,$MeV and $f=93\,$MeV)
 and experimental \cite{Hagiwara:pw}
 branching ratios. We shall refer to these results as those with ``universal
 couplings''. 

\begin{table}[htbp]
\begin{center}
\begin{tabular}{|c||c||c|c|}\hline  
$i$ &$C_i$  &$B_i^{th}$ & $B_i^{exp}$ \\ \hline \hline  
 $\rho\pi^0\gamma$  & $\sqrt{\frac{2}{3}}$  & $7.1\times 10^{-4}$ &$(7.9\pm 2.0)\times 10^{-4}$ \\ \hline  
 $\rho\eta\gamma$ & $\frac{2}{\sqrt{3}}$  &  $5.7\times 10^{-4}$   &  $(3.8\pm 0.7)\times 10^{-4}$ \\ \hline  
 $\omega\pi^0\gamma$ &$\sqrt{2}$  & $12.0$\%  &  $8.7\pm 0.4$\% \\ \hline  
$\omega\eta\gamma$ & $\frac{2}{3\sqrt{3}}$ & $12.9\times 10^{-4}$  & $(6.5\pm 1.1)\times 10^{-4}$ \\ \hline  
 $\phi\eta\gamma$& $\frac{2}{3}\sqrt{\frac{2}{3}}$ & $0.94$\% &
  $1.24\pm 0.10$\% \\ \hline  
 $\phi\pi^0\gamma$ &0 &-- &-- \\ \hline  
${ {K^{*+} \to K^+\gamma} \atop {K^{*-} \to K^-\gamma}}$ 
&$\frac{\sqrt{2}}{3}(2-\frac{M_{\omega}}{M_{\phi}})$  & $13.3\times 10^{-4}$  &  $(9.9\pm 0.9)\times 10^{-4}$ \\ \hline  
 ${ {K^{*0}\to K^{0}\gamma} \atop 
 {\overline K^{*0}\to \overline K^{0}\gamma} }$
 &$-\frac{\sqrt{2}}{3}(1+\frac{M_{\omega}}{M_{\phi}})$ & $27.3\times 10^{-4}$ & $(23\pm 2)\times 10^{-4}$ \\   
\hline
\end{tabular}
\caption{   SU(3) $C_i$ coefficients together with
 theoretical and experimental branching ratios for different 
vector meson decay processes. }
\end{center}
 \end{table}

The agreement with the data is fair but the
results can be improved by incorporating $SU(3)$ 
breaking mechanisms \cite{Bramon:1994pq}. For that purpose,
we will
normalize here the $C_i$ couplings so that the 
branching ratios in Table 1 agree with experiment.
These will be called results with ``normalized couplings''.
In this way we are taking into account 
phenomenologically the corrections to the $VP\gamma$
vertex from an underlying field theory.

Once the $VP\gamma$ couplings have been fixed,  we can use
them in the VMD amplitude corresponding to the diagram
of Fig.~1, which is given by
\begin{eqnarray}
-it^{VMD}&=& \{ i\sqrt{6}\frac{1}{q^2-M_{\rho}^2+iM_{\rho}\Gamma(q^2)}
\left( G\frac{2}{3} e \frac{G_V}{M_\rho} \right) ^2
\cdot
\left\vert \begin{array}{ccc}
q\cdot q & q\cdot k_2  & q\cdot \epsilon_2\\
k_1\cdot q & k_1\cdot k_2 & k_1\cdot \epsilon_2\\
\epsilon_1\cdot q & \epsilon_1\cdot k_2 & \epsilon_1\cdot
\epsilon_2
\nonumber
\end{array} \right\vert \\
&& + (k_1\leftrightarrow k_2,\,q\to q') \} 
+ \{ \rho \to \omega
\}, 
\label{tvmd}
\end{eqnarray} 
where $q=P-k_1$, $q'=P-k_2$, with $P,k_1,k_2$ the momentum of the
$\eta$ and the two photons. We have parametrized
the $\rho$ width phenomenologically as: 
$\Gamma_\rho(q,s)=\frac{(6.14)^2}{48\pi s}(s^2-4 s\,m_\pi^2)^{3/2}$
whereas for the $\omega$ we have considered a constant
$\Gamma_\omega=8.44\,$ MeV. Nevertheless, our results are rather 
insensitive to these details. From the above amplitude,
the $\eta$ decay width is easily calculated, as well as the
$\gamma\gamma$ invariant mass distribution
\begin{equation}
\frac{d\Gamma}{dM_I}=\frac{1}{16(2\pi)^4m_{\eta}^2}M_I
\int_0^{m_\eta-\omega}dk_1\int_0^{2\pi}d\phi\ \, \Theta(1-A^2)
\, \Sigma\mid t\mid^{2},
\end{equation} 
where we take for reference the momentum of the pion,
$\vec{p}$, in the $z$ direction, so that
\begin{eqnarray}
\vec{p} = p \, \left(
\begin{array}{c}
0\\
0\\
1\\
\end{array} \right),
\quad
\vec{k_1} = k_1 \left(
\begin{array}{c}
sin\,\theta \ \cos\,\phi \\
sin\,\theta\ sin\,\phi\\
cos\,\theta\\
\end{array} \right),
\quad
\vec{k_2}=-(\vec{k_1}+\vec{p}),\hspace{2cm}\\
p=\frac{\lambda^{1/2}(m_{\eta}^2,M_I^2,m_{\pi}^2)}{2m_{\eta}},\quad
A\equiv
cos(\gamma_1\pi^0)=\frac{1}{2k_1p}[(m_{\eta}-
\omega-k_1)^2-k_1^2-\vert\vec{p}\vert^2]
\end{eqnarray}
with $\omega$ the energy of the $\pi^0$.

In Fig.~2 we show the results of the mass distribution with
and without the radiative widths normalization factors.
The integrated width is given by
$\Gamma=0.57\,$eV (universal couplings); 
$\Gamma=0.30 \pm 0.06\,$eV (normalized couplings),
where the error has been calculated from a Monte Carlo Gaussian sampling
of the normalization parameters within the errors of the experimental 
branching ratios of Table 1. Let us note that there have been 
stable values for the vector meson radiative  
widths throughout the last decade in the PDG but a sizable change in the
PDG2002. Had we used the PDG2000, we would have obtained
$0.21 \pm 0.05\,$eV (normalized couplings). 

It is interesting to compare these results with those in
\cite{Ametller:1991dp}, where they used a universal SU(3) coupling
with $G_V$
adjusted to the $\omega\to \pi^0\gamma$ decay data existing at that time,
and obtained an "all orders" value of $0.31\,$eV. 
The difference between that value and the $0.21 \pm 0.05\,$eV that
we would have obtained with older data has to be attributed
to the adjusting to all the
branching ratios, instead of just one as in \cite{Ametller:1991dp}.

\begin{figure}[h]
\vspace{.3cm}
\centerline{\hbox{\psfig{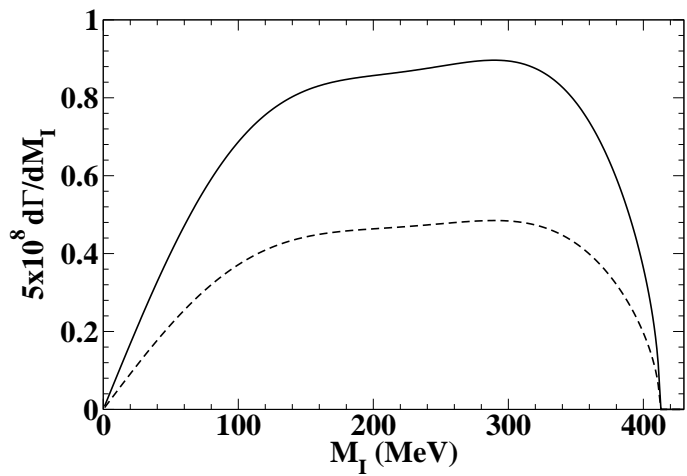}}}
\label{fig:fig2}
\caption{\rm 
Invariant mass distribution of the two photons 
with VMD terms only. The solid curve
has been calculated with an universal coupling, whereas the dashed
one  has the couplings  normalized differently
to fit the radiative decays.}
\end{figure}

Our  VMD normalized
 result is within three standard deviations from the value presently given
in \cite{exp},\cite{Hagiwara:pw}:  $\Gamma=0.84 \pm 0.18\,$eV,
but within one sigma of the more recent one 
presented in \cite{nefkens}, $\Gamma=0.42 \pm 0.14\,$eV.
There are, however, other contributions that we consider next.

\section{Meson loops}

The contribution of pion loops to $\eta \to \pi^0 \gamma \gamma$,
evaluated  in \cite{Ametller:1991dp},
proceeds, to begin with, through the  G-parity violating $\eta\to
\pi^0\pi^+\pi^-$ process.  Since the contribution
 is  proportional to $m_u-m_d$, 
it is very small and we 
think that if such terms
are included, other isospin violating terms proportional 
to $m_u-m_d$, and isospin
violating corrections to the main terms should also be included. 
Rather than undertaking
this delicate task, we will  use the results of \cite{Ametller:1991dp} 
to estimate uncertainties
from all these sources. 

The main meson loop contribution comes from  the charged kaon 
 loops, calculated at  $O(p^4)$ 
in
\cite{Ametller:1991dp,Bel'kov:1995fj,Bellucci:1995ay,Bijnens:1995vg,Ko:1992zr,Jetter:1995js},
 and proceeds via $\eta \to \pi^0 K^+ K^-\rightarrow \pi^0\gamma\gamma$.
Note that these loops are also suppressed due to the large kaon masses.
That is why the 
$\eta\rightarrow\pi^0 a_0(980)\rightarrow\pi^0\gamma\gamma$ mechanism 
was included explicitly, with uncertainties
in the size and sign of the $a_0(980)$ couplings.
As commented in the introduction, the chiral unitary approach
solves this problem by generating dynamically the $a_0(980)$
in the $K^+ K^-\rightarrow \pi^0\eta$ amplitude.

In this section, we will illustrate this approach 
by revisiting the work done in \cite{Oller:1997yg} on the
related process $\gamma\gamma\to
\pi^0\eta$ where  the chiral unitary approach was successfully
applied around the $a_0(980)$ region. 
Since for the $\eta$ decay the low energy region of
$\gamma\gamma\to
\pi^0\eta$ is also of interest, we will include
next the VMD mechanisms also in this reaction. 
Once we check that we describe correctly
$\gamma\gamma\to
\pi^0\eta$, the results can be easily translated
to the eta decay. We will finally add other 
anomalous meson loops that are numerically relevant for eta decay but not 
for $\gamma\gamma\to
\pi^0\eta$.

\subsection{The unitarized $\gamma\gamma\to\pi^0\eta$ 
amplitude in the $a_0(980)$ region}

In \cite{Oller:1997yg} it was shown that,
within the unitary chiral approach, the $\gamma\gamma\to\pi^0\eta$ amplitude 
around the $a_0(980)$ region, diagrammatically represented at one
loop in Fig.~3,
factorizes as
\begin{equation}
-it=(\tilde{t}_{\chi K}+ \tilde{t}_{AK^+K^-}   )t_{K^+K^-,\pi^0\eta}
\label{eq:tNPA629}
\end{equation}
with $t_{K^+K^-,\pi^0\eta}$ the full $K^+K^-\to\pi^0\eta$ transition
amplitude.

\begin{figure}[h]
\centerline{\hbox{\psfig{file=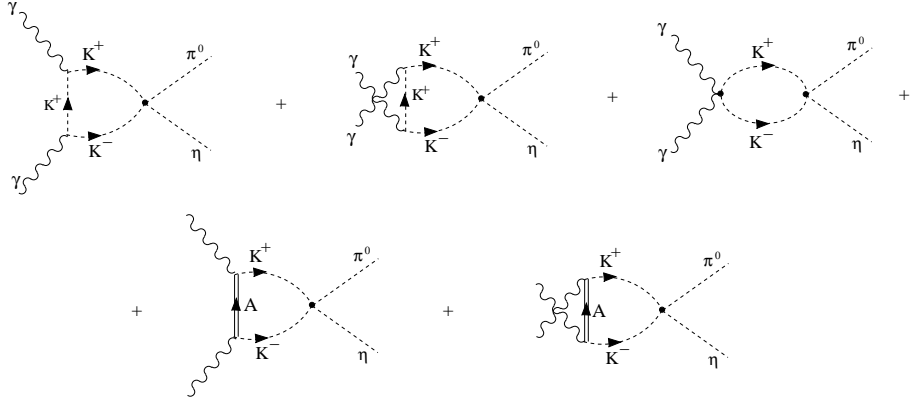,width=12cm}}}
\caption{\rm
Diagrams for the chiral loop contribution  }
\end{figure}
The first three diagrams correspond to $\tilde{t}_{\chi K}\,
t_{K^+K^-,\pi^0\eta}$ of Eq.~(\ref{eq:tNPA629}), already evaluated 
at one loop in 
\cite{Bijnens:1987dc,Donoghue:ee}, where the factorization 
of the leading $t_{K^+K^-,\pi^0\eta}$ also occurred. In
our case $\tilde{t}_{\chi K}$, written in a general
 gauge to be also used for the $\eta\to
\pi^0 \gamma\gamma$ reaction,  is given by
\begin{equation}
\tilde{t}_{\chi K} = - \frac{2 e^2 }{16 \pi^2}
\left( g^{\mu\nu}-\frac{k_{2\mu}k_{1\mu}}
{ k_1 \cdot k_2 } \right) \epsilon_{1\mu}\epsilon_{2\mu}
\,
\left\{ 1 + \frac{m_K^2}{s} \, \left[ \log \left( \frac{1 + (1 - 4 m_K^2 
/ s)^{\frac{1}{2}}}
{1 - (1 - 4 m_K^2/s)^{\frac{1}{2}}} \right
) - i \pi\right]^2 \right\},
\end{equation}
above the $K^+K^-$ threshold, with the $-i\pi$ term removed below
threshold. Note that 
 the unitarized $t_{K^+K^-,\pi^0\eta}$
transition matrix, not just the lowest order chiral amplitude,
is factorized outside the loop integral. This on shell
factorization was shown in \cite{Oller:1997yg}
 by proving  that the off shell
part of the meson-meson amplitude did not contribute to the
loop integral.

The meson meson scattering amplitude was evaluated in
\cite{Oller:1997ti} by summing
the Bethe Salpeter (BS) equation with a kernel formed from the
lowest order meson chiral Lagrangian amplitude and
regularizing the loop function with a three momentum cut off.
Subsequently, other approaches like the inverse amplitude
method \cite{OllOsePel,GomezNicola:2001as} or the $N/D$ method
\cite{Oller:1999zr} were used and all of them
gave the same results in the meson scalar sector. 
For $\gamma\gamma\to\pi^0\eta$ below 1 GeV
only the 
$L=0,\,I=1$ amplitudes are needed \cite{L2}. 
The BS equation sums the
diagrammatic series of Fig.~4,
\begin{figure}[h]
\centerline{\hbox{\psfig{file=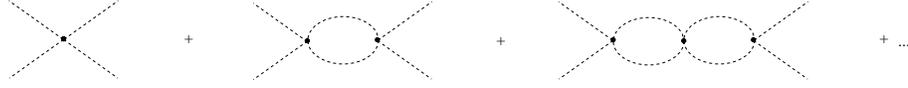,width=12cm}}}
\caption{\rm
Diagrams summed in the BS equation, using the $O(p^2)$ ChPT vertices.  }
\end{figure}
which implies that in the $\gamma\gamma\to\pi^0\eta$ transition
of Fig.~3 one is resumming the diagrams of Fig.~5.
\begin{figure}[h]
\centerline{\hbox{\psfig{file=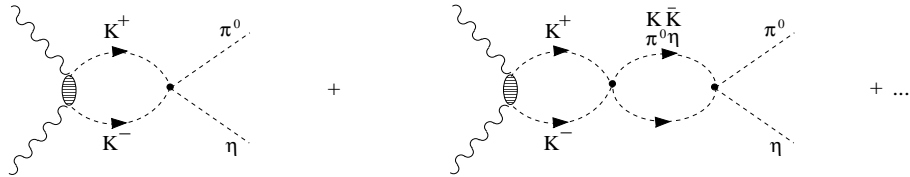,width=12cm}}}
\caption{\rm
Resummation for $\gamma\gamma\rightarrow\pi^0\eta$.}
\end{figure}
Furthermore, the same on-shell factorization of the $t$ matrix
in the loops found for $\gamma\gamma\to\pi^0\eta$
was also justified for meson-meson scattering in
\cite{Oller:1997ti}.  Thus, the BS equation with coupled channels can be 
 solved algebraically, leading to the following solution in matrix form 
\begin{equation}
t(s)=[1-t_2(s)G(s)]^{-1}t_2(s),
\end{equation}
with $s$ the invariant mass of the two mesons, $t_2$ the lowest
order chiral amplitude and $G(s)$ a diagonal matrix,
$\mbox{diag}(G_{\overline{K}K},G_{\eta\pi})$, accounting for the loop
functions of two mesons. These $G$ functions were regularized in
\cite{Oller:1997ti}
by 
means of a cut off.  The  $G$ analytic expressions, both 
using a cut off or dimensional regularization 
can be found in \cite{Oller:1999zr}.

In Eq.~(\ref{eq:tNPA629}) there is another term, $\tilde{t}_{AK^+K^-}
t_{K^+K^-,\pi^0\eta}$, which corresponds to the last two
diagrams of Fig.~3 where the axial vector meson $K_1(1270)$
 is exchanged. For
the one loop result we follow 
\cite{Donoghue:1993kw}. Given the large mass of the axial
vector, both the factorization of the unitarized 
on shell meson-meson scattering
amplitude outside the loop, as well as that of the $\gamma\gamma\to
K^+K^-$ amplitude are also justified
\cite{Oller:1997yg}. Hence, when the full series of 
Fig.~5 is considered,  one obtains the contribution
$\tilde{t}_{AK^+K^-}t_{K^+K^-,\pi^0\eta}$ with $\tilde{t}_{AK^+K^-}$ given by
\begin{equation}
\hspace*{-1.1cm}
\tilde{t}_{AK^+K^-} = - 2 e^2 \left( g^{\mu\nu}-\frac{k_{2\mu}k_{1\mu}}
{ k_1 \cdot k_2 } \right) \epsilon_{1\mu}\epsilon_{2\mu} 
\frac{\displaystyle{ (L^r_9 + L^r_{10})}}{\displaystyle{f^2}}
\left[ \frac{\displaystyle{s_A}}{\displaystyle{2 \beta (s)}} l n 
\left( \frac{\displaystyle{1 + \beta (s) + \frac{s_A}{s}}}
{\displaystyle{1 - \beta (s) + \frac{s_A}{s}}}
\right) + s\right]\cdot G_{K\overline{K}},
\end{equation}
with $s_A=2(m_A^2-m_K^2)$, and 
$\beta(s)=(1-\frac{4m_K^2}{s})^{1/2}$.\\

First of all we show in Fig.~6 the result for the 
$\gamma\gamma\to\pi^0\eta$ cross section obtained from Eq.~(\ref{eq:tNPA629}),
 which
coincides with that obtained in \cite{Oller:1997yg}.
\begin{figure}[h]
\centerline{\hbox{\psfig{file=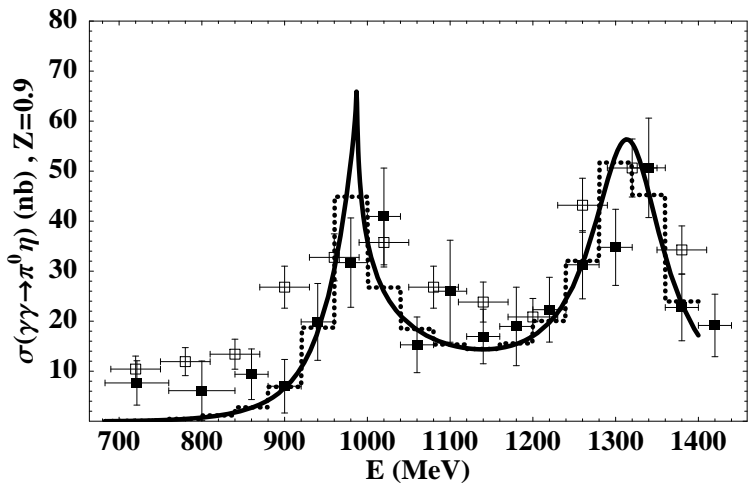,width=10cm}}}
\caption{\rm
$\gamma\gamma\rightarrow\pi^0\eta$ cross section, using Eq.~(\ref{eq:tNPA629}).
Z is the maximum value of $cos(\theta)$ integrated.
The experimental data come from \cite{oest,antre}, 
the latter ones normalized in the
$a_2(1320)$ peak region. 
The dashed histogram corresponds to the convolution
over an experimental resolution of 40 MeV.}
\label{fig:ggexp}
\end{figure}
To ease the comparison with experimental data we also show
 the events concentrated in bins of 40 MeV, roughly like the
experimental ones.
 We can easily notice the peak of the $a_0(980)$
whose dynamical generation
is guaranteed by the resummation of diagrams in Fig.~5.
The resummed 
$t_{K^+K^-\rightarrow\pi^0\eta}$ amplitude has indeed a
pole in the complex plane associated to
the $a_0(980)$ resonance  \cite{Oller:1997ti}.

In Fig.~6, we also show results above and below $a_0(980)$, 
whose description requires further ingredients 
than those needed just for the $a_0(980)$ region.
In particular,
the $a_2(1320)$ resonance 
(second peak) is included phenomenologically
as in ref.~\cite{Oller:1997yg}.

In \cite{Oller:1997yg} loops like those in Fig.~3,
but exchanging a vector meson 
instead of an axial-vector meson were estimated negligible
in the $a_0(980)$ region and hence neglected.
In addition, the VMD tree level mechanism of Fig.~1 (with an outgoing $\eta$)
was neglected since it has no resonant structure
 in the $\gamma\gamma$
s-channel.
As a consequence the agreement of Eq.~(\ref{eq:tNPA629}) 
with experiment is fair 
but some discrepancies can be noticed in Fig.~6
at low energies. 

\subsection{VMD mechanisms in $\gamma\gamma\to\eta\pi^0$}

For the purpose of the present work, 
$\eta\to \pi^0\gamma\gamma$, the low energy region of
the $\gamma\gamma\to\pi^0\eta$ reaction is
also relevant. Therefore we will include as a novelty 
both the VMD tree level contribution as well as the 
loops involving vector meson exchange. 

First, we can see in Fig.~7 that the results obtained adding the tree level VMD 
amplitude normalized to the $\omega, \rho$ radiative decay rates
(dashed line) are acceptable
around and below the peak of the  $a_0(980)$ resonance.
Let us note that the inclusion of these terms does improve the 
description of the low energy region.
The binning of the theoretical results would make again
the apparent agreement with data to look much better, but for the sake of
clarity we have not added more lines to the figure, as long as
the binning effect has already been illustrated 
in Fig.~6.
Although in section one we have justified the use of the normalized
couplings, we also show 
in Fig.~7 the result using universal couplings (dotted line).
In this process, the
effect of normalizing the
couplings of the vector meson radiative decays is not as drastic as
in 
Fig.~2 for the $\eta$ decay where only the VMD mechanism was considered.
In what follows we will only use the normalized couplings.

\begin{figure}[h]
\centerline{\hbox{\psfig{file=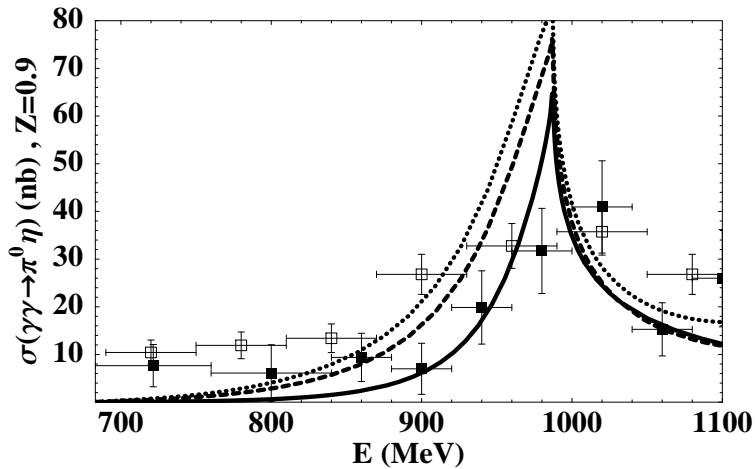,width=10cm}}}
\caption{\rm
$\gamma\gamma\rightarrow\pi^0\eta$ cross section, using
Eq.~(\ref{eq:tNPA629} (continuous line), adding the universal VMD contribution (dotted line) or
the normalized VMD contribution (dashed line).}
\end{figure}

Second,  in addition to the axial vector meson
exchange in loops considered in the previous section,
we have to include the loops with vector meson exchange for completeness.
In fact, some of the uncertainties estimated in 
\cite{Ametller:1991dp} were linked to these loops. 
For consistency, once again we have to sum
the series obtained by iterating the loops in the four meson
vertex shown in Fig.~8.
Hence, the new amplitude, which we shall call $t^{VMDL}$, will be given by:
\begin{equation}
  t^{VMDL}=t_{\eta\pi^0,\eta\pi^0}(M_I)G_{\eta\pi}
\tilde t^{VMD}_{\eta\pi}(M_I)
\left[ \epsilon_1\epsilon_2-\frac{(k_2\epsilon_1)
(k_1\epsilon_2)}{k_1 k_2}\right]
\label{VMDloops1}
\end{equation}
where now $ \tilde t^{VMD}_{\eta\pi}$ is the factor that multiplies
the $\epsilon_1\epsilon_2$ product in the $s$-wave projection of the
$ t^{VMD}_{\eta\pi}$ amplitude in the $\gamma\gamma\rightarrow\pi^0\eta$ CM. 
Although the Lorentz structure
of polarization vector products may seem rather complicated
from Eq.~(\ref{tvmd}), it is easy to show that after the s-wave projection
the polarization vectors factorize indeed as $\epsilon_1\epsilon_2$.
In a general frame
the $\epsilon_1\epsilon_2$ factor has to be replaced by 
$\epsilon_1\epsilon_2-(k_2\epsilon_1)(k_1\epsilon_2)/(k_1 k_2)$.
Once again we have factorized the amplitudes for the same reasons
 as done with the other terms.

\begin{figure}[h]
\centerline{\hbox{\psfig{file=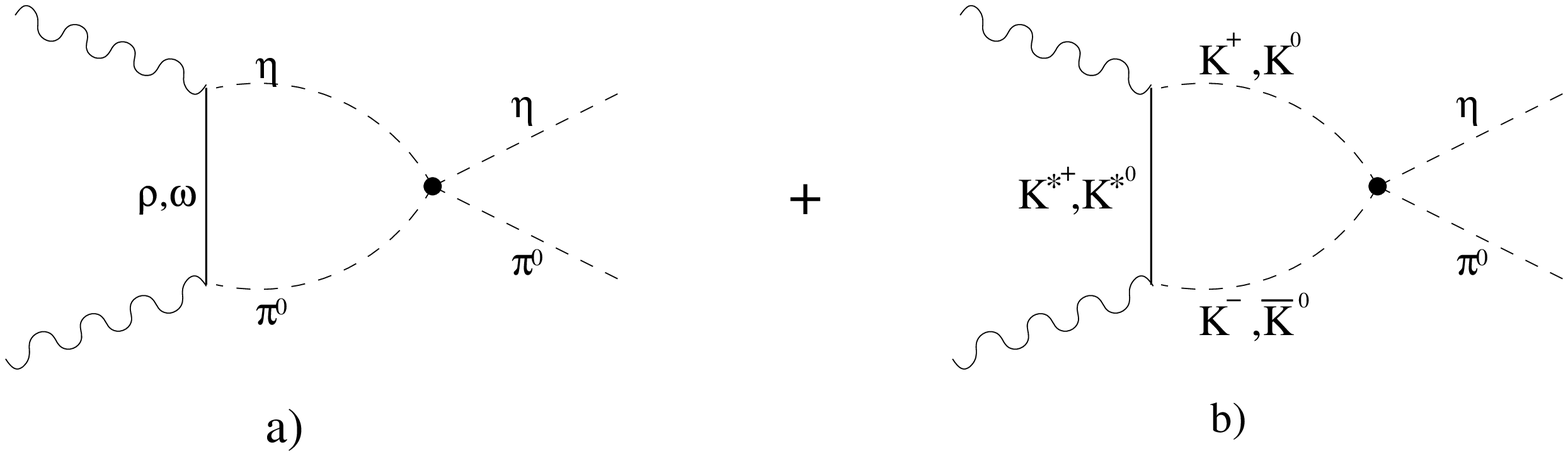,width=14cm}}}
\caption{\rm
Loop diagrams for VMD terms. The diagrams with the two crossed
photons are not depicted but are also included in the calculations.}
\end{figure}
Of course, when introducing loops with vector meson exchange
we have to consider loops involving 
a $K^{*+}$
or a $K^{*0}$ exchanged between the photons (see Fig.~8.b),
which were not present at tree level. 
These would be given by
\begin{eqnarray}
  \label{VMDloops2}
  t^{VMDL}_{K\bar{K}}&=&\left(t_{\eta\pi^0,K^+K^-}(M_I)G_{K\bar{K}}(M_I)
\tilde T^{VMD}_{K^+K^-}(M_I)\right.
\\ 
&&+\left.
t_{\eta\pi^0,K^0\bar{K}^0}(M_I)G_{K\bar{K}}(M_I)
\tilde T^{VMD}_{K^0\bar{K}^0}(M_I)\right)\left[ \epsilon_1\epsilon_2-\frac{(k_2\epsilon_1)(k_1\epsilon_2)}{k_1 k_2}\right] \nonumber
\end{eqnarray}

The contribution of all these new VMD loop diagrams 
is an increase of the order of 10-20\% of the result
shown in Fig.~7 by the dashed line (normalized VMD couplings). 
The new result would overlap in a large region with the dotted line
of Fig.~7 and hence we do not show it explicitly.

In what follows we make some considerations about 
the diagrammatic interpretation of the ``all order'' 
VMD calculation, the normalization of the $VP\gamma$
vertices and the meson-meson interaction. By ``all order'' 
VMD one means \cite{Ametller:1991dp} that the full vector meson
propagator $(s-M_V^2+iM_V\Gamma(s))^{-1}$ is used in the calculations.
This full propagator includes self-energy diagrams 
in a Dyson-Schwinger resummation, leading to a shift of the bare mass
and generating a width \cite{Oller:2000ug}. Thus, one must think
in terms of self-energy insertions in the middle of
the vector meson lines in Fig.~1 and 8.
The  $VP\gamma$ coupling normalization  to agree with the radiative
vector meson decays can also be understood as considering
 vertex correction diagrams in Fig.1 and 8, 
and therefore it does not lead to any double counting with the dressing of
the vector meson propagator.
Finally, the meson-meson interaction in the VMD terms 
leads to the diagrams of Fig.~8, in which the two 
pseudoscalar mesons interact
through four-pseudoscalar meson vertices.
The resummation of pseudoscalar meson-meson loops
thus leaves apart the vector meson lines and the 
$VP\gamma$ vertices. Once again this ensures that there is no double
counting.

\subsection{Meson loops in $\eta\to\pi^0\gamma\gamma$}

In the $\eta\to\pi^0\gamma\gamma$ case, the meson loop diagrams
correspond to those of $\pi^0\eta\to\gamma\gamma$ but
considering the $\pi^0$ as an outgoing particle.
Hence, it is enough to replace $s=(p_{\eta}+p_{\pi})^2$
by $M_I^2=(p_{\eta}-p_{\pi})^2=(p_{\gamma_1}+p_{\gamma_2})^2$
in all the  $\pi^0\eta\to \gamma\gamma$ amplitudes,
which factorize in all the loop diagrams that we have considered
so far, and in the $\tilde{t}_{\chi K}$ and $\tilde{t}_{A K^+K^-}$ function.

Since we are considering
all the VMD diagrams and the chiral loops, 
we still have to take into account
another kind of loop diagrams
\cite{Ametller:1991dp}, shown in Fig.~9, which involve two anomalous 
$\gamma\rightarrow 3 M$ vertices.
Despite being  $O(p^8)$ 
it has been  found \cite{Ametller:1991dp} that they can have a
 non negligible effect on the $\eta$ decay.
The further rescattering of the mesons in the diagrams 
of Fig.~9, given the structure of
the $\gamma MMM$ vertex \cite{Ametller:1991dp} in the momenta of the
particles, would be
suppressed by factors of $\vec{p}\,^2_{\gamma}/\vec{q}\,^2$ 
(with $q$ the loop variable) with
respect to those considered for the VMD mechanism. 
This, and the fact that these anomalous terms are
already small,
 makes the consideration of  
rescattering in these loops superfluous.
Therefore, it is enough to 
take the results
from \cite{Ametller:1991dp} where it is found that their largest contribution
comes from the kaon loops. 
We use Eqs.(12),(13) and (27) of that reference (note that there is a 
global change of sign
with respect to our notation).
Concerning $\gamma\gamma\rightarrow\eta\pi^0$,
these kind of loops have been neglected in the previous section, 
because the intermediate particles are very far off shell, due to
 the crossed character of the loop in the
reaction.

\begin{figure}[h]
\centerline{\hbox{\psfig{file=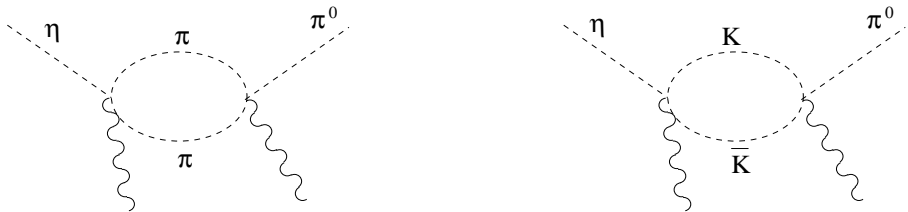,width=12cm}}}
\caption{\rm 
Diagrams with two anomalous $\gamma\rightarrow 3 M$ vertices.}
\end{figure}

\section{Results for $\eta\to\pi^0\gamma\gamma$}

Using the model described in the previous sections, we
plot in Fig.~10 the different contributions to $d\Gamma/dM_I$.
We can see that the largest contribution is that of the 
tree level VMD
(long dashed line). 
Let us recall that this is a new result as long as 
we are using the VMD couplings normalized to 
agree with the vector radiative decays.
The resummation of the loops in Fig.~3 using
Eq.(\ref{eq:tNPA629}),
(short dashed line) gives a small
contribution ($0.011\,$eV in the total width), 
but when added coherently to the tree level VMD, 
leads to an increase of $30\,$\% in the $\eta$ decay
rate (dashed-dotted line). More interestingly, the shape of the $\gamma\gamma$ 
invariant mass distribution is appreciably changed with respect to
the tree level VMD, developing a peak at
high invariant masses. The 
resummed VMD loops in Fig.~8, using Eqs.(\ref{VMDloops1}) 
and (\ref{VMDloops2}), leads, through interference,
to a moderate increase of the $\eta$ decay rate (double dashed
dotted line), smaller 
than that of the chiral loops considered before. The last
ingredient is the contribution of the anomalous
mechanisms of Fig.~9 (continuous line), 
leading again to a moderate increase of
the $\eta$ decay rate, also smaller than 
the chiral loops from Eq.(\ref{eq:tNPA629}).
These anomalous mechanisms have a very similar shape to the tree level VMD
and interfere with it in the whole range of invariant masses.

%\vspace{0.9cm}
\begin{figure}[h]
\centerline{\hbox{\psfig{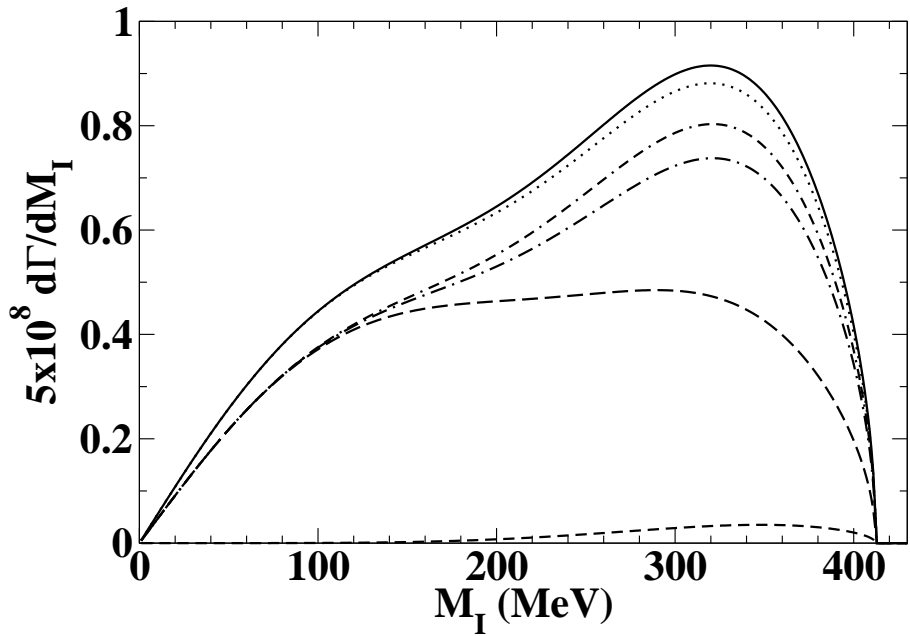}}}
\label{fig:fig11}
\caption{\rm 
Contributions to the two photon invariant mass distribution.
 From bottom to top,
short dashed line: chiral loops from Eq.(\ref{eq:tNPA629}); 
long dashed line: only tree level VMD;
dashed-dotted line:
coherent sum of the previous mechanisms; double
dashed-dotted line: idem  but adding the resummed VMD loops;
continuous line: idem but adding the anomalous terms of Fig.~10,
which is the full model presented in this work 
(we are also showing as a dotted line the
full model but substituting the full 
$t_{K^+K^-,\eta\pi^0}$
amplitude by its lowest order).
}
\end{figure}
 
Altogether, when integrating over the invariant mass, we get:
\begin{equation}
  \Gamma(\eta\rightarrow\pi^0\gamma\gamma)=0.47\pm 0.08 \,\mbox{eV}.
\end{equation}
Note that the inclusion of the loops has increased the
tree level VMD result by 50\%. 
For comparison, we quote  here what
we would have obtained using 
the universal VMD couplings: $0.80\,$eV.

So far, the theoretical error 
has been obtained only from 
the propagation of the experimental errors
in the vector meson radiative decay branching
ratios, given in Table~1.
The errors from this source had not been considered before
although they will turn out to produce the largest uncertainty.
In practice we generate a Gaussian weighted 
random value for each VMD coupling which
yield a result for the width. This procedure is
repeated a sufficiently large number of times, 
leading to an approximate
Gaussian distribution
of results from where we obtain a central value and the error.

We come now to revisit the uncertainties considered in  
\cite{Ametller:1991dp}. 
One of the largest sources to the $\pm0.2\,$eV
accepted uncertainties in that work, was the contribution of 
the $a_0(980)$, whose sign was unknown.
This problem is solved in the present work
since the $a_0(980)$ is generated dynamically from the 
rescattering of the mesons implied in the Bethe-Salpeter 
resummation of the $t_{K^+K^-,\eta\pi^0}$ amplitude.
Hence its effect can be easily observed 
comparing with the standard ChPT result, which is obtained 
by substituting the full  $t_{K^+K^-,\eta\pi^0}$
by its lowest order $O(p^2)$. In Fig.~10 this corresponds to
the difference between the continuous and the dotted line.
The contribution of the $a_0(980)$ resonance tail is rather small and increases
the $\eta$ decay rate from $0.47$eV to $0.48\,$eV. The sign of its
contribution is unambiguously determined. Thus, the present calculation removes
completely this source of error. The explicit calculation of the $a_0(980)$ 
contribution giving such a small effect justifies neglecting the
$a_2(1320)$ resonance contribution which lies much further away in energy. 

The other source of uncertainty in  \cite{Ametller:1991dp}
was the contribution of the VMD loops. 
We have been able to calculate them in this work and, as seen in Fig.~10,
these effects are also rather small. They increase the $\eta$ decay rate by
$0.02\,$eV. Altogether the $a_0(980)$ plus  the VMD loops
increase the $\eta$ decay
rate by $0.03\,$eV. We thus eliminate these two sources of previous
uncertainties,  while realizing at the same time that
the uncertainties of $0.2\,$eV attributed to these sources in 
\cite{Ametller:1991dp} were indeed a generous upper bound.

In our approach the tree level exchange
of the $h_1(1170)$, $b_1(1235)$  and $h_1(1380)$
axial
resonances \cite{Ko:1992zr}
will be included as an uncertainty. The reason is that,
according to 
 \cite{Ko:1992zr}, they
would increase the decay width by about $0.07\,$eV.
However, as shown in  \cite{Ko:1992zr,Jetter:1995js}, their inclusion in 
$\gamma\gamma\rightarrow\pi^0\pi^0$ with the couplings used in 
 \cite{Ko:1992zr} would overestimate the $\gamma\gamma\rightarrow\pi^0\pi^0$
cross section. In view of these discrepancies, 
we thus consider safe to 
accept a theoretical uncertainty 
of the order of $0.05\,$eV which
should still be a
generous upper bound. 

As commented at the beginning of section three, 
there are uncertainties due to isospin violating terms.
We estimate the errors from this source using the results
obtained in \cite{Ametller:1991dp} for the G-parity violating term
corresponding to Fig.~3 but with pion loops.
This contribution is of the order of 
$0.05\,$eV to the total $\eta$ decay rate.

Finally, by summing all errors in quadrature, we arrive to
\begin{equation}
\Gamma(\eta\to\pi^0\gamma\gamma)=0.47\pm 0.10\,eV
\label{resultfinal}
\end{equation}

Note that although we have considered a new error source 
from the uncertainties in the vector radiative decays,
which turns out to be the largest one,
we still have reduced the uncertainty from previous calculations.

The result of Eq.~(\ref{resultfinal}) is in remarkable agreement 
with the latest experimental number \cite{nefkens},
and lie within two sigmas from the 
 earlier ones in \cite{exp,Hagiwara:pw}. Confirmation
of those preliminary results  would therefore be important to test
the consistency of this new approach. Furthermore,
precise measurements of  
the $\gamma\gamma$ invariant mass distributions would 
be of much help given the differences found
 with and without loop contributions.

\section{Conclusions}

We have reanalyzed the $\eta\rightarrow\pi^0\gamma\gamma$
decay within the context of meson chiral lagrangians,
gathering all the mechanisms discussed in the literature,
but improving them in the following aspects:

On the one hand, using the well
tested chiral unitary approach,
we have removed the uncertainties
from the $a_0(980)$ resonance as well as those from loops
with the exchange of one vector meson. In particular, 
since the $a_0(980)$ is generated dynamically from the meson loop
resummation,
we have unambiguously
determined the sign of its  contribution, whereas for the
one vector loops we have performed an explicit calculation 
that in previous works had only been considered as a 
large source of uncertainty.

On the other hand, we have also checked the consistency
with other related processes: First,  a relevant observation
is that the tree level vector meson dominance
 amplitude with a universal SU(3) vector-vector-pseudoscalar 
coupling  is not consistent with the
$\rho\rightarrow\eta\gamma$, $\omega\rightarrow\pi^0\gamma$ and 
$\omega\rightarrow\eta\gamma$ decays. Consequently, 
throughout the  $\eta\rightarrow\pi^0\gamma\gamma$ calculation,
we have used couplings normalized to agree with the 
radiative vector meson decays.
Second, we have established the consistency of our 
$\eta\rightarrow\pi^0\gamma\gamma$ model with the related 
$\gamma\gamma\to\eta\pi^0$ process.

Furthermore, we have performed a careful
error analysis of our results. As a novelty we have considered the
experimental errors in the vector meson radiative decay widths,
which turn out to be the largest source of
uncertainty. However, since, as just commented above,
 we have removed former sources of uncertainty, our
final error is still smaller than previous estimates.

Altogether we have found a result of
$\Gamma(\eta\to\pi^0\gamma\gamma)=0.47\pm0.10\,$eV.  

With the improved calculation just presented, it seems
clear that the mechanisms thus far suggested in the literature
in the context of meson chiral lagrangians lead to a result
at variance with the experimental result
$\Gamma=0.84 \pm 0.18\,$eV from \cite{exp,Hagiwara:pw}.
However, it is worth noticing the agreement
of the above result with
the new preliminary measurement $\Gamma=0.42 \pm 0.14\,$eV
 from \cite{nefkens}.
Nevertheless, a measurement of the invariant mass distribution would be
more stringent. Confirmation of the preliminary results of  \cite{nefkens}
and an accurate measurement of the $\gamma\gamma$ 
invariant mass distribution should 
be the experimental priorities 
to clarify the situation.

\section*{Acknowledgments}
We are specially grateful to J.A. Oller for fruitful discussions, 
technical help and  his
careful reading of the manuscript.
We would also like to acknowledge useful discussions with J. Bijnens.
One of us, L.R., acknowledges support from the Consejo Superior de
Investigaciones Cient\'{\i}ficas. J.R.P. acknowledges financial
support from a CICYT-INFN collaboration grant as well as a
Marie Curie fellowship MCFI-2001-01155. He also thanks the
Dipartimento de Fisica, Universita' de Firenze-INFN Sezione di Firenze
for its hospitality. 
This work is also
partly supported by DGICYT contract numbers BFM2000-1326, PB98-0782, and the 
E.U. EURIDICE network contract no. HPRN-CT-2002-00311.


\begin{thebibliography}{99}

\footnotesize

\bibitem{exp} D. Alde et al., Yad. Fiz 40 (1984) 1447; D. Alde et al., Z. Phys.
C25 (1984) 225;  L.G. Landsberg, Phys. Rep. 128 (1985) 301.

\bibitem{Hagiwara:pw}
K.~Hagiwara {\it et al.}  [Particle Data Group Collaboration],
%``Review Of Particle Physics,''
Phys.\ Rev.\ D {\bf 66} (2002) 010001.
%%CITATION = PHRVA,D66,010001;%%

%\cite{Ng:sc}
\bibitem{Ng:sc}
J.~N.~Ng and D.~J.~Peters,
%``A Study Of Eta $\to$ Pi0 Gamma Gamma Decay Using The Quark Box Diagram,''
Phys.\ Rev.\ D {\bf 47} (1993) 4939.
%%CITATION = PHRVA,D47,4939;%%

%\cite{Nemoto:1996bh}
\bibitem{Nemoto:1996bh}
Y.~Nemoto, M.~Oka and M.~Takizawa,
%``$\eta \to \pi~0 \gamma \gamma$ decay in the three-flavor Nambu-Jona-Lasinio model,''
Phys.\ Rev.\ D {\bf 54} (1996) 6777
[arXiv:hep-ph/9602253].
%%CITATION = HEP-PH 9602253;%%

%\cite{Ametller:1991dp}
\bibitem{Ametller:1991dp}
L.~Ametller, J.~Bijnens, A.~Bramon and F.~Cornet,
%``Chiral perturbation theory for $\eta \to \pi~0 \gamma \gamma $ ,''
Phys.\ Lett.\ B {\bf 276} (1992) 185.
%%CITATION = PHLTA,B276,185;%%

\bibitem{oneda} S. Oneda and G. Oppo, Phys. Rev. 160 (1968) 1397.

%\cite{Picciotto:sn}
\bibitem{Picciotto:sn}
C.~Picciotto,
%``Rate And Spectrum Of Eta $\to$ Pi0 Gamma Gamma,''
Nuovo Cim.\ A {\bf 105} (1992) 27.
%%CITATION = NUCIA,A105,27;%%

%\cite{Bel'kov:1995fj}
\bibitem{Bel'kov:1995fj}
A.~A.~Bel'kov, A.~V.~Lanyov and S.~Scherer,
%``gamma gamma $\to$ pi0 pi0 and eta $\to$ pi0 gamma gamma at O(p**6) in the  NJL model,''
J.\ Phys.\ G {\bf 22} (1996) 1383
[arXiv:hep-ph/9506406].
%%CITATION = HEP-PH 9506406;%%

%\cite{Bellucci:1995ay}
\bibitem{Bellucci:1995ay}
S.~Bellucci and C.~Bruno,
%``Gamma gamma $\to$ pi0 pi0 and eta $\to$ pi0 gamma gamma at low-energy within the extended Nambu-Jona-Lasinio model,''
Nucl.\ Phys.\ B {\bf 452} (1995) 626
[arXiv:hep-ph/9502243].
%%CITATION = HEP-PH 9502243;%%

%\cite{Bijnens:1995vg}
\bibitem{Bijnens:1995vg}
J.~Bijnens, A.~Fayyazuddin and J.~Prades,
%``The $\gamma\gamma\to\pi~0\pi~0$ and $\eta\to\pi~0\gamma\gamma$ Transitions in the Extended NJL Model,''
Phys.\ Lett.\ B {\bf 379} (1996) 209
[arXiv:hep-ph/9512374].
%%CITATION = HEP-PH 9512374;%%

%\cite{Achasov:2001qm}
\bibitem{Achasov:2001qm}
M.~N.~Achasov {\it et al.},
%``Search for the radiative decay eta $\to$ pi0 gamma gamma in the SND  experiment at VEPP-2M,''
Nucl.\ Phys.\ B {\bf 600} (2001) 3
[arXiv:hep-ex/0101043].
%%CITATION = HEP-EX 0101043;%%

%\cite{Ko:1992zr}
\bibitem{Ko:1992zr}
P.~Ko,
%``Contributions to the C odd axial vector resonances to eta $\to$ pi0 gamma gamma and gamma gamma $\to$ pi0 pi0,''
Phys.\ Rev.\ D {\bf 47} (1993) 3933.
%%CITATION = PHRVA,D47,3933;%%
%\cite{Ko:rg}

\bibitem{Ko:rg}
P.~Ko,
%``Eta $\to$ Pi0 Gamma Gamma And Gamma Gamma $\to$ Pi0 Pi0 In O(P**6) Chiral Perturbation Theory,''
Phys.\ Lett.\ B {\bf 349} (1995) 555
[arXiv:hep-ph/9503253].
%%CITATION = HEP-PH 9503253;%%

%\cite{Jetter:1995js}
\bibitem{Jetter:1995js}
M.~Jetter,
%``Eta $\to$ pi0 gamma gamma to O (p**6) in chiral perturbation theory,''
Nucl.\ Phys.\ B {\bf 459} (1996) 283
[arXiv:hep-ph/9508407].
%%CITATION = HEP-PH 9508407;%%

%\cite{Oest:1990ki}
\bibitem{Oest:1990ki}
T.~Oest {\it et al.}  [JADE Collaboration],
%``Measurement Of Resonance Productions In The Reactions Gamma Gamma $\to$ Pi0 Pi0 And Gamma Gamma $\to$ Pi0 Eta,''
Z.\ Phys.\ C {\bf 47} (1990) 343.
%%CITATION = ZEPYA,C47,343;%%

%\cite{Antreasyan:1985wx}
\bibitem{Antreasyan:1985wx}
D.~Antreasyan {\it et al.}  [Crystal Ball Collaboration],
%``Formation Of Delta (980) And A2 (1320) In Photon-Photon Collisions,''
Phys.\ Rev.\ D {\bf 33} (1986) 1847.
%%CITATION = PHRVA,D33,1847;%%


%\cite{Oller:1997ti}
\bibitem{Oller:1997ti}
J.~A.~Oller and E.~Oset,
%``Chiral symmetry amplitudes in the S-wave isoscalar and isovector  channels and the sigma, f0(980), a0(980) scalar mesons,''
Nucl.\ Phys.\ A {\bf 620}, 438 (1997)
[Erratum-ibid.\ A {\bf 652}, 407 (1997)].
%%CITATION = HEP-PH 9702314;%%

%\cite{Kaiser:fi}
\bibitem{Kaiser:fi}
N.~Kaiser,
%``Pi Pi S-Wave Phase Shifts And Non-Perturbative Chiral Approach,''
Eur.\ Phys.\ J.\ A {\bf 3} (1998) 307.
%%CITATION = EPHJA,A3,307;%%

\bibitem{OllOsePel}
 J.A. Oller, E. Oset and J. R. Pel\'aez, Phys. Rev. Lett. \textbf{80} (1998) 3452;
 Phys. Rev. D \textbf{59} (1999) 74001; Erratum-ibid.D60 (1990) 099906 
 
\bibitem{Oller:1999zr}
J.~A.~Oller and E.~Oset,
%``N/D description of two meson amplitudes and chiral symmetry,''
Phys.\ Rev.\ D {\bf 60} (1999) 074023.
%%CITATION = HEP-PH 9809337;%%

%\cite{Nieves:2000bx}
\bibitem{Nieves:2000bx}
J.~Nieves and E.~Ruiz Arriola,
%``Bethe-Salpeter approach for unitarized chiral perturbation theory,''
Nucl.\ Phys.\ A {\bf 679} (2000) 57
[arXiv:hep-ph/9907469].
%%CITATION = HEP-PH 9907469;%%


%\cite{GomezNicola:2001as}
\bibitem{GomezNicola:2001as}
A.~Gomez Nicola and J.~R.~Pelaez,
%``Meson meson scattering within one loop chiral perturbation theory and  its unitarization,''
Phys.\ Rev.\ D {\bf 65} (2002) 054009
[arXiv:hep-ph/0109056].
%%CITATION = HEP-PH 0109056;%%


\bibitem{Oller:1997yg}
J.~A.~Oller and E.~Oset,
%``Theoretical study of the gamma gamma $\to$ meson meson reaction,''
Nucl.\ Phys.\ A {\bf 629} (1998) 739
[arXiv:hep-ph/9706487].
%%CITATION = HEP-PH 9706487;%%


%\cite{Dobado:1992zs}
\bibitem{Dobado:1992zs}
A.~Dobado and J.~R.~Pelaez,
%``Unitarity and gamma gamma $\to$ pi pi in chiral perturbation theory,''
Z.\ Phys.\ C {\bf 57} (1993) 501.
%%CITATION = ZEPYA,C57,501;%%


\bibitem{Yamagishi:1995kr}
H.~Yamagishi and I.~Zahed,
%``A Master formula for chiral symmetry breaking,''
Annals Phys.\  {\bf 247} (1996) 292
[arXiv:hep-ph/9503413].
%%CITATION = HEP-PH 9503413;%%

%\cite{Lee:1998mz}
\bibitem{Lee:1998mz}
C.~H.~Lee, H.~Yamagishi and I.~Zahed,
%``Master formulae approach to photon fusion reactions,''
Nucl.\ Phys.\ A {\bf 653} (1999) 185
[arXiv:hep-ph/9806447].
%%CITATION = HEP-PH 9806447;%%

%\cite{Bramon:2001un}
\bibitem{Bramon:2001un}
A.~Bramon, R.~Escribano, J.~L.~Lucio Martinez and M.~Napsuciale,
%``Scalar sigma meson effects in rho and omega decays into pi0 pi0 gamma,''
Phys.\ Lett.\ B {\bf 517} (2001) 345
[arXiv:hep-ph/0105179].
%%CITATION = HEP-PH 0105179;%%

%\cite{Palomar:2001vg}
\bibitem{Palomar:2001vg}
J.~E.~Palomar, S.~Hirenzaki and E.~Oset,
%``Chiral loops and VMD in the V $\to$ P P gamma decays,''
Nucl.\ Phys.\ A {\bf 707} (2002) 161
%%CITATION = HEP-PH 0111308;%%

%\cite{Marco:1999df}
\bibitem{Marco:1999df}
E.~Marco, S.~Hirenzaki, E.~Oset and H.~Toki,
%``Radiative decay of rho0 and Phi mesons in a chiral unitary approach,''
Phys.\ Lett.\ B {\bf 470} (1999) 20
[arXiv:hep-ph/9903217].
%%CITATION = HEP-PH 9903217;%%

\bibitem{nefkens} 
S.~Prakhov, in Proceedings of the International Conference of Non-Accelerator New Physics, Dubna,
Russia; 
B.~M.~Nefkens and J.~W.~Price, in the Eta Physics Handbook, Proc. of the
Uppsala Workshop, http://www.tsl.uu.se/~faldt/eta/Proceedings.html,
Published in Phys.\ Scripta {\bf T99} (2002) 114
[arXiv:nucl-ex/0202008].
%%CITATION = NUCL-EX 0202008;%%

%\cite{Gasser:1983yg}
\bibitem{Gasser:1983yg}
J.~Gasser and H.~Leutwyler,
%``Chiral Perturbation Theory To One Loop,''
Annals Phys.\  {\bf 158} (1984) 142.
%%CITATION = APNYA,158,142;%%

%\cite{Ecker:1988te}
\bibitem{Ecker:1988te}
G.~Ecker, J.~Gasser, A.~Pich and E.~de Rafael,
%``The Role Of Resonances In Chiral Perturbation Theory,''
Nucl.\ Phys.\ B {\bf 321} (1989) 311.
%%CITATION = NUPHA,B321,311;%%

%\cite{Bramon:1992kr}
\bibitem{Bramon:1992kr}
A.~Bramon, A.~Grau and G.~Pancheri,
%``Intermediate vector meson contributions to V0 $\to$ P0 P0 gamma decays,''
Phys.\ Lett.\ B {\bf 283} (1992) 416.
%%CITATION = PHLTA,B283,416;%%


%\cite{Bramon:1994pq}
\bibitem{Bramon:1994pq}
A.~Bramon, A.~Grau and G.~Pancheri,
%``Radiative vector meson decays in SU(3) broken effective chiral Lagrangians,''
Phys.\ Lett.\ B {\bf 344} (1995) 240.
%%CITATION = PHLTA,B344,240;%%
%\cite{Oller:1997yg}

%\cite{Bijnens:1987dc}
\bibitem{Bijnens:1987dc}
J.~Bijnens and F.~Cornet,
%``Two Pion Production In Photon-Photon Collisions,''
Nucl.\ Phys.\ B {\bf 296} (1988) 557.
%%CITATION = NUPHA,B296,557;%%

%\cite{Donoghue:ee}
\bibitem{Donoghue:ee}
J.~F.~Donoghue, B.~R.~Holstein and Y.~C.~Lin,
%``The Reaction Gamma Gamma $\to$ Pi0 Pi0 And Chiral Loops,''
Phys.\ Rev.\ D {\bf 37} (1988) 2423.
%%CITATION = PHRVA,D37,2423;%%

\bibitem{L2} H. Albrecht et al. Z. Phys.{\bf C48} (1989) 183.

%\cite{Donoghue:1993kw}
\bibitem{Donoghue:1993kw}
J.~F.~Donoghue and B.~R.~Holstein,
%``Photon-photon scattering, pion polarizability and chiral symmetry,''
Phys.\ Rev.\ D {\bf 48} (1993) 137
[arXiv:hep-ph/9302203].
%%CITATION = HEP-PH 9302203;%%

\bibitem{oest} T.Oest et al., Z. Phys. C47 (1990) 343.

\bibitem{antre} D. Antreasyan et al., Phys. Rev. D33 (1986) 1847.

%\cite{Oller:2000ug}
\bibitem{Oller:2000ug}
J.~A.~Oller, E.~Oset and J.~E.~Palomar,
%``Pion and kaon vector form factors,''
Phys.\ Rev.\ D {\bf 63} (2001) 114009
[arXiv:hep-ph/0011096].
%%CITATION = HEP-PH 0011096;%%
  
 
\end{thebibliography}
\end{document}